%% file: main.tex
\documentclass[conference]{IEEEtran}
\usepackage{cite}
\usepackage{graphicx}
\graphicspath{ {images/} }

\newtheorem{definition}{Definition}[section]

\usepackage{subfigure}

\usepackage{times}

\usepackage[T1]{fontenc}
\usepackage[scaled=0.81]{beramono}

\newcommand{\e}[1]{\mbox{\lstinline|#1|}}

\usepackage{amssymb}
\usepackage{amsmath}
\usepackage{graphicx}
\usepackage{epstopdf}
\usepackage{colortbl}
\usepackage[counterclockwise, figuresleft]{rotating}

\usepackage{listings}
\usepackage{eiffel}
\usepackage{boogie}
\usepackage{framed}
\usepackage{fancyvrb}
\usepackage{caption}
\usepackage{epigraph}

\usepackage[utf8]{inputenc} 
\usepackage[T1]{fontenc} 

\usepackage{zed-csp}
\usepackage{circle}

\usepackage[autostyle=true]{csquotes} 

\begin{document}


\title{Seamless Object-Oriented Requirements}


\author
{\IEEEauthorblockN{Alexandr Naumchev}
\IEEEauthorblockA{Software Engineering Lab\\
Innopolis University\\
Innopolis, Russia\\
}
}
\maketitle


\begin{abstract}
Design by Contract enables seamless software development by unifying software requirements with their implementations.
In its pure form, however, Design by Contract leaves some problems with contracts' expressiveness, verifiability, and reusability open.
These problems significantly reduce practical applicability of seamless development.
The present article introduces seamless object-oriented requirements -- a novel approach to seamless development that builds upon Design by Contract and now-available advanced program proving tools.
The article explains and illustrates the new approach, concluding with a quantitative evaluation of the extent to which the approach fixes the problems of traditional contracts.
\end{abstract}

\begin {IEEEkeywords}
seamless object-oriented requirements,
seamless development,
Design by Contract,
Eiffel,
AutoProof,
object-oriented programming,
program proving, formal verification
\end{IEEEkeywords}

\section{Introduction}
\label{intro} 
\input{sections/Introduction.tex}

\section{The Unified Solution}
\label{unified_solution}
\input{sections/TheUnifiedSolution.tex}




\section{Quantitative Evaluation}
\label{quantitative_evaluation}
\input{sections/QuantitativeEvaluation.tex}

\input{sections/Summary.tex}

\bibliographystyle{IEEEtran}
\bibliography{extracted}

\end{document}

%% file: sections/Introduction.tex
Design by Contract (DbC \cite{DBLP:journals/computer/Meyer92}) made it possible to:
\begin{itemize}
    \item Specify software requirements directly inside programs.
    \item Check correctness of the programs directly against the specified requirements, both through testing \cite{DBLP:journals/computer/MeyerFCLWS09} and program proving \cite{DBLP:conf/tacas/TschannenFNP15}.
\end{itemize}
\begin{figure*}[h]
    \fbox{\includegraphics[width=0.95\textwidth]{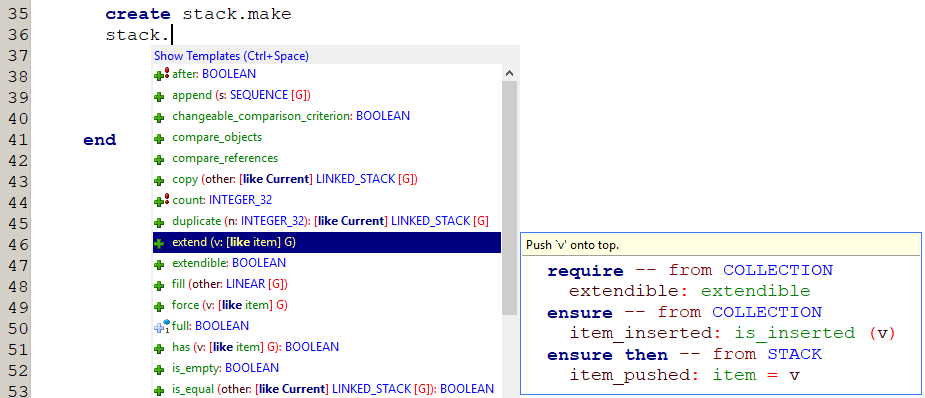}}
    \caption{EiffelStudio displaying hints, including contracts and natural language comments.}
    \label{fig:sota:eiffel_studio}
\end{figure*}
Contracts are irreplaceable in how they document software components.
Figure \ref{fig:sota:eiffel_studio} depicts EiffelStudio during the programming process.
More concretely, it depicts a situation in which the programmer has just entered a dot symbol after a variable and is looking for a feature to call.
EiffelStudio offers the list of features callable on the variable.
Going through the list causes the selected feature's documentation to appear in the rightmost pop-up window.
It contains the natural language description of the feature along with its semantics in the form pre- and postconditions.
The ability to see the callable features' meanings may significantly speed-up the programming process.

DbC, while offering powerful mechanisms for specifying and verifying object-oriented programs, fails to capture their formal properties of certain forms:
\begin{itemize}
    \item Multicommand abstract data type (ADT \cite{DBLP:journals/acta/GuttagH78}) axioms, such as \e{pop (push (stack, element)) = stack}.
    Non-trivial ADT specifications always include multicommand axioms \cite{DBLP:books/daglib/0025377}, and a program specification mechanism should be in place for checking them.
    \item Temporal properties, such as $\square (Stimulus \implies \Diamond Response)$.
    Some temporal properties of software were found to follow certain patterns \cite{DBLP:conf/icse/DwyerAC99}, \cite{DBLP:conf/icse/KonradC05}
    Some of these patterns have complicated structures, making their application a non-trivial task.
\end{itemize}
Addressing these problems requires an alternative specification approach that would, at the same time, inherit the best properties of contracts. 

In addition to the generally insufficient expressiveness, contracts may be weak.
A program may be correct against a weak contract and, at the same time, include incorrect instructions that are not ruled out by the weak contract.
Consider the following Eiffel function for computing the square of the input number:
\begin{lstlisting}
square (x: INTEGER): INTEGER
  do
    -- Computing the square of x.
  ensure
    non_negative_result: Result >= 0
  end
\end{lstlisting}
Clearly, the result should be non-negative.
The \e{non_negative_result} assertion, however, leaves room for incorrect implementations: there are infinitely many non-negative functions.
A well-defined contract unambiguously defines all the side-effects and return values of the enclosing procedure.
Specifying a contract's well-definedness as part of the contract itself seems to be unrealistic.

Contracts also miss support for reuse of recurring specification patterns across conceptually unrelated data structures.
Consider the following procedure from a stack implementation:
\begin{lstlisting}
push (element: G)
  do
    -- Pushing `element' onto the current stack.
  ensure
    increments_count: count = old count + 1
  end
\end{lstlisting}
Consider also the following procedure from a turnstile controller implementation:
\begin{lstlisting}
insert_coin
  do
    -- Handling coin insertion.
  ensure
    increments_coins: coins = old coins + 1
  end
\end{lstlisting}
The \e{increments_count} and \e{increments_coins} assertions follow the same common pattern: both increment the value of a class attribute.
Technically, DbC makes it possible to reuse contracts from ancestor classes.
Turnstiles, however, conceptually have nothing in common with stacks, and inheriting them from a common class would violate the object-oriented principles \cite{DBLP:books/ph/Meyer97}.
DbC offers no mechanism other than inheritance to reuse contracts with repeated fundamental semantics.

%% file: sections/TheUnifiedSolution.tex
Our approach relies on the following fundamental principles to fix the problems with contracts:
\begin{itemize}
    \item Use the imperative subset of the programming language for specifications, not only the declarative contracts.
    \item Use loops equipped with loop invariants and variants for specifying temporal and timing properties.
    \item Use Hoare logic based static program proving to infer correct contracts from axiomatically correct implementations.
    \item Use object-oriented genericity and abstraction to make software requirements universally reusable.
\end{itemize}
Methodologically, object-oriented software construction \cite{DBLP:books/ph/Meyer97} becomes the requirements specification method, and DbC \cite{DBLP:journals/computer/Meyer92} empowered with the static program prover becomes the requirements verification method.

\subsection{The choice of notation and technology} \label{unified_solution:notation}

We chose Eiffel as the representative language to illustrate our concepts.
It has human-friendly syntax, natively supports contracts and builds around object-oriented concepts.
An advanced technology stack accompanies Eiffel.
AutoProof, the contracts-based modular program prover \cite{DBLP:conf/tacas/TschannenFNP15}, allowed us to work at the cutting edge of the programming technology.
It has been playing a key role in our studies.
AutoProof is a program prover based on Hoare logic \cite{DBLP:journals/cacm/Hoare69} extended with \emph{semantic collaboration} \cite{DBLP:conf/fm/PolikarpovaTFM14} -- reasoning framework that covers phenomena specific to object-oriented programming, such as aliasing, callbacks and information hiding.
Polikarpova et al. demonstrated practical applicability of AutoProof by using it to fully verify EiffelBase2 -- a specified library of containers \cite{DBLP:journals/fac/PolikarpovaTF18}.
We have been using EiffelBase2 extensively as a valuable source of data for testing our ideas.

\subsection{Imperative instructions for software requirements} \label{unified_solution:specification_drivers}

Design by Contract \cite{DBLP:journals/computer/Meyer92} was originally designed under the assumption that the contracts would be checked at run time.
Practitioners were perceiving code solely as an executable artifact.
AutoProof makes it possible to use program elements as statically verifiable statements that may never be executed.
This possibility has been the main thinking vehicle driving the development of the thesis.

Specification drivers operationalize this possibility.
The article describes several novel concepts, among which the notion of specification driver is the most fundamental.
Understanding this concept is essential for understanding the rest of the work: the other concepts build on top of specification drivers.
Syntactically, a specification driver is an object-oriented Hoare triple, or a self-contained contracted routine.
The following specification driver formally captures the multicommand stack axiom mentioned in Section \ref{intro}:
\begin{lstlisting}
push_then_pop (s_1, s_2: STACK [G]; x: G)
-- pop (push (s, x)) = s
-- Popping a stack after pushing an element on it
-- results in the original stack, assuming that
-- these operations only modify the stack itself.
  require
    s_1 ~ s_2
  modify
  	s_1
  do
    s_1.push (x)
    s_1.pop
  ensure
    s_1 ~ s_2
  end
\end{lstlisting}
The natural language comment captures the axiom's mathematical representation and informal description.
The \e{push_then_pop} routine depends only on its formal parameter and is self-contained in that sense.
The routine may be submitted for static verification to AutoProof.
AutoProof can only rely on contracts of the supplier classes for verification.
Hence, it will only accept the \e{push_then_pop} routine if the contracts of \e{push} and \e{pop} are correct.
We treat the \e{push_then_pop} routine, as well as any other specification driver, as an axiomatically correct code, and infer correct contracts from these axiomatic definitions.
The \e{modify} clause captures the frame condition, critical for static verification.
The \e{require} and \e{ensure} clauses capture the routine's pre- and postcondition, respectively.

\begin{definition}
	A \textbf{specification driver} is a self-contained contracted routine that captures some behavioral property of its formal parameters through the contract.
\end{definition}

A separate work \cite{DBLP:conf/tase/NaumchevM16} gives more intuition behind specification drivers and how to apply them in the presence of a program prover like AutoProof.
The subsequent sections develop this idea further and find for it more complex applications -- way more complex than specification and verification of stack.
Specification drivers:
\begin{itemize}
	\item Capture temporal properties and timing constraints in addition to ADT axioms.
	\item Capture contracts' well-definedness axioms for checking with AutoProof.
	\item Capture repeated requirements' formal semantics in a reusable form.
\end{itemize}
The remaining sections expand, detail and illustrate these benefits of specification drivers.
The article concludes with the generalized object-oriented treatment of requirements with specification drivers serving as the verification mechanism.
They became the main thinking vehicle taking us to the general notions of seamless object-oriented requirement (SOOR) and SOOR template (SOORT).

\subsection{Loops for temporal properties} \label{unified_solution:temporal_properties}

While ADT axioms are good at capturing properties of software components, requirements to control software are mainly specified as temporal properties \cite{DBLP:conf/icse/DwyerAC99}.
Specification drivers appeared to be applicable not only to capturing multicommand ADT axioms, but also to capturing temporal properties.
The \e{popping_empties_stack} procedure below formally specifies the temporal property written as a natural language comment inside the procedure:
\begin{lstlisting}
  popping_empties_stack (s: V_STACK [INTEGER])
  -- Repeatedly popping a non-empty stack will
  -- make it empty within s.count steps.
    require
      stimulus: not s.is_empty
    modify
      s
    do
      from
      until
        s.is_empty
      loop
        s.pop
      variant
        s.count
      end
    end
\end{lstlisting}
The \e{require} precondition clause characterizes the situation to which the requirement applies: the stack must not be empty.
The \e{until} loop exit condition expresses the desired property of the requirement.
The \e{variant} clause limits the maximum number of iterations to \e{s.count}, according to the limitation stated in the requirement.
The loop body between the \e{loop} and the \e{variant} keywords applies the \e{pop} operation constrained by the requirement.
Applying the Eiffel plus AutoProof combination for specifying and verifying temporal properties resulted in finding a major flaw in a published executable specification of a landing gear system \cite{AutoReq}.

\subsection{Genericity and abstraction for reusable requirements} \label{unified_solution:reusable_specification_patterns}

Dwyer et al. \cite{DBLP:conf/icse/DwyerAC99} analyzed 555 temporal control software requirements and successfully mapped 511 of them to only 23 temporal SRPs.
The \e{popping_empties_stack} example from Section \ref{unified_solution:temporal_properties} represents an example of the most widespread temporal SRP, which covered 241 requirements out of the 555 analyzed in \cite{DBLP:conf/icse/DwyerAC99}.
Specification drivers, together with the standard object-oriented genericity mechanism, successfully capture such repeated SRPs.
The \e{STIMULUS_RESPONSE} deferred class below captures the popular SRP followed by the \e{popping_empties_stack} example:
\begin{lstlisting}
deferred class STIMULUS_RESPONSE [S]
-- Deferred definitions start here.
  stimulus (s: S): BOOLEAN
    deferred
    end
  response (s: S): BOOLEAN
    deferred
    end
  verification_timer (s: S): INTEGER
    deferred
    end    
  action (s: S)
    deferred
    end    
-- Deferred definitions end here.

-- Reusable semantics starts here.
  verify (s: S)
    require
      stimulus (s)
    modify
      s
    do
      from
      until
        response (s)
      loop
        action (s)
      variant
        verification_timer (s)
      end
    ensure
      response (s)
    end
-- Reusable semantics ends here.
end    
\end{lstlisting}
We call such classes \emph{seamless object-oriented requirement templates (or SOORTs)}.
The \e{POPPING_EMPTIES_STACK} class below instantiates the \e{STIMULUS_RESPONSE} SOORT and captures the same property as the \e{popping_empties_stack} specification driver from Section \ref{unified_solution:temporal_properties}:
\newpage
\begin{lstlisting}
class POPPING_EMPTIES_STACK [G]
inherit STIMULUS_RESPONSE [STACK [G]]
  stimulus (s: STACK [G]]): BOOLEAN
    do
      Result := not s.is_empty
    end
  response (s: STACK [G]]): BOOLEAN
    do
      Result := s.is_empty
    end
  verification_timer (s: STACK [G]]): INTEGER
    do
      Result := s.count
    end    
  action (s: STACK [G]])
    do
      s.pop
    end    
end    
\end{lstlisting}
We call such classes \emph{seamless object-oriented requirements (or SOORs)}.
Two main features make specifying temporal properties through SOOR(T)s better than doing it through plain specification drivers:
\begin{itemize}
    \item SOORTs apply to arbitrary data types.
    \item SOORs do not need to explicitly state the potentially tricky temporal semantics; they simply inherit it from the parent SOORT in the form of the \e{verify} specification driver. 
\end{itemize}
We have successfully specified all the 23 recurring temporal SRPs in the form of SOORTs \cite{DBLP:conf/tools/Naumchev19}.
While the complexity of the original temporal SRPs varies significantly, the complexity of SOORs inheriting from our SOORTs remains at the same level.
In a recent work \cite{DBLP:conf/tools/Naumchev19} we polish the notions of SOORs and SOORTs to the level at which the specification of resulting SOORs relies purely on inheritance.

\subsection{Well-definedness of contracts} \label{unified_solution:well_definedness}

Specification drivers enable proof-based reasoning about well-definedness of individual features.
The \e{push_is_well_defined} procedure below specifies the well-definedness property for the \e{push} procedure from type \e{STACK [G]}, where \e{G} is a generic type parameter coming from the enclosing class' declaration:
\begin{lstlisting}
push_is_well_defined (s1, s2: STACK [G]; x: G)
-- Whenever two stack instances are equal, they should
-- remain equal after being called in identical ways.
  require
    s1.is_equal (s2)
  modify
    s1, s2
  do
    s1.push (x)
    s2.push (x)
  ensure
    s1.is_equal (s2)
  end
\end{lstlisting}
The program prover, AutoProof in our case, will only accept the \e{push_is_well_defined} procedure if the contract of the \e{push} procedure is well-defined with respect to the definition of equality, \e{is_equal}.
We separately handle the nuances of specifying the equivalence relation in the original work devoted to the early idea of specification drivers \cite{DBLP:conf/tase/NaumchevM16}.
Applying well-definedness specification drivers to EiffelBase2, a fully verified container library \cite{DBLP:journals/fac/PolikarpovaTF18}, uncovered numerous features with weak contracts.
Section \ref{quantitative_evaluation} details these findings.

\subsection{Key activities} \label{unified_solution:activities}

Several major activities characterize our approach.

\subsubsection{Developing a SOORT} \label{unified_solution:activities:soort_development}

Developing a SOORT requires the same skills as developing any other object-oriented class.
It assumes identification of an SRP, hardcoding its immutable part and parameterizing its variable part through abstraction and genericity.
\begin{enumerate}
	\item Identify the SRP's formal semantics.
	\item Declare the SOORT class and name it to reflect the identified semantics.
	\item Encode the identified semantics through specification drivers and put them inside the SOORT class.
	\item Make the specification drivers work with generic, not actual types; make the generic types part of the enclosing SOORT's declaration.
\end{enumerate}
A separate work \cite{DBLP:conf/tools/Naumchev19} provides more technical details to help developing new SOORTs.

\subsubsection{Instantiating a SOORT into a SOOR} \label{unified_solution:activities:soor_specification}

Converting an input requirement to a SOOR assumes identifying an SRP that cover the requirement, inheriting from the SOORT capturing the identified SRP, and implementing the SOORT's deferred definitions parts.
The resulting SOOR class must be fully defined.
\begin{enumerate}
	\item Find the SOORT encoding the identified semantics.
	\item Create a concrete class inheriting from the found SOORT.
	\item Replace the SOORT's formal generic parameters with actual generic parameters.
	\item Implement the SOORT's deferred features.
	\item Make sure that the newly implemented SOOR successfully compiles.
\end{enumerate}

\subsubsection{Verifying through program proving} \label{unified_solution:activities:verify_through_proving}

Proving correctness of a candidate implementation against a SOOR consists of running AutoProof on the SOOR.
In this case, AutoProof will check correctness of the SOOR's specification drivers against the candidate solution's contracts.
This may require writing additional annotations on the specification drivers that capture the SOOR's formal semantics.
\begin{enumerate}
	\item Run AutoProof on the SOOR.
	\item If AutoProof rejects the input, fix the implementation contract; go to step 1.
	\item If AutoProof accepts the input, consider the implementation contract correct.
	\item Implement the derived contract and check the implementation's correctness with AutoProof.
\end{enumerate}

%% file: sections/QuantitativeEvaluation.tex
The present section discusses quantitative arguments showing that SOORs promote expressiveness, contracts' well-definedness, and reusability.

\subsection{Expressiveness} \label{quantitative_evaluation:expressiveness}

The evidence of the SOORTs' expressiveness comes from the possibility to capture:

\begin{itemize}
	\item The 23 temporal SRPs for control software \cite{DBLP:conf/tools/Naumchev19}.
	\item The repeated real-time semantics \cite{DBLP:conf/icse/KonradC05}, through the standard notion of loop variants.
	\item The 21 ADT specifications recurring in the requirements literature \cite{DBLP:conf/tools/Naumchev19}.
\end{itemize}

Some of the control software SRPs have tricky formal semantics.
For example, the ``Bounded Existence Between Q and R'' SRP, where the bound is at most 2 designated states, looks as follows in LTL:
\begin{equation}
	\begin{aligned}
		\square ((Q \land & \Diamond R) \implies ((\lnot P \land \lnot R) \mathcal{U} (R \lor ((P \land \lnot R) \mathcal{U} (R \lor\\
		& ((\lnot P \land \lnot R) \mathcal{U} (R \lor ((P \land \lnot R) \mathcal{U} (R \lor (\lnot P \mathcal{U} R)))))
	\end{aligned}
	\label{eq:bounded_existence_between}
\end{equation}
We were able to encode this formula as the \e{verify} specification driver inside the \e{BOUNDED_EXISTENCE_BETWEEN} class.
Moreover, representing requirements' formal semantics as specification drivers allows us to generalize from the 2-states to the k-states case.
Three out of the five notations used by Dwyer et al.
-- LTL, CTL and GIL -- lack expressiveness for performing such generalization.
Using the programming language as a requirements notation makes it possible to perform the generalization through enclosing the bounded existence semantics into an additional loop that runs exactly k times.

\subsection{Well-definedness of contracts} \label{quantitative_evaluation:verifiability}

We have evaluated the usefulness of specification drivers for analyzing contracts' well-definedness.
The EiffelBase2 library \cite{DBLP:journals/fac/PolikarpovaTF18} seems to be a perfect data set for such evaluation -- its authors claim well-definedness of all the contracts in this library.
We analyzed well-definedness of feature \e{copy_} in the EiffelBase2 classes.
The feature copies the given object into the current one.
Out of the 17 versions of the feature, 6 were underspecified.
They come from the following classes:
\begin{itemize}
	\item \e{V_ARRAY2}
	\item \e{V_LINKED_QUEUE}
	\item \e{V_LINKED_STACK}
	\item \e{V_ARRAYED_LIST_ITERATOR}
	\item \e{V_ARRAY_ITERATOR}
	\item \e{V_HASH_SET_ITERATOR}
\end{itemize}
Deeper analysis revealed that the most common problem was not taking into consideration the possibility of aliasing between the copied and the current objects.
For the \e{V_HASH_SET_ITERATOR} class, however, AutoProof did not accept the well-definedness axiom even with the aliasing prohibited in the precondition.
AutoProof did not terminate when checking the well-definedness axiom for the following 2 classes:
\begin{itemize}
	\item \e{V_DOUBLY_LINKED_LIST_ITERATOR}
	\item \e{V_LINKED_LIST_ITERATOR}
\end{itemize}
The non-termination may be interpreted as if the features were underdefined.
Summarizing the results of the analysis, out of the 17 versions AutoProof accepted the well-definedness axiom only for 9.
Underdefined contracts may have security implications.
Consider appending the following code to the implementation of feature \e{copy_} in class \e{V_ARRAY2}:
\begin{center}
	\begin{lstlisting}
	else
	  array.wipe_out
	  row_count := 0
	  column_count := 0
	end
	\end{lstlisting}
\end{center}
The \e{else} clause describes the aliasing situation, which is ignored in the contract of the feature.
The added code wipes out the current array's data.
AutoProof accepts the modified implementation, which is not what we expect: a feature responsible for copying from another array should not erase the current one.
The published presentation of EiffelBase2 claims well-definedness of the flawed classes \cite{DBLP:journals/fac/PolikarpovaTF18}.

The EiffelBase2 library contains software components.
As for control software, expressing their properties as specification drivers was also fruitful.
A separate work \cite{AutoReq} details uncovering an error in a published abstract state machine (ASM) implementation \cite{DBLP:journals/sttt/ArcainiGR17} of the Landing Gear System (LGS) \cite{boniol2014landing} -- a commonly used example for evaluating applicability of formal specification and verification techniques.
								
\subsection{Reusability} \label{quantitative_evaluation:reusability}

We might evaluate the extent to which our approach improves reusability by following a commonly accepted practice -- measuring the amount of duplication removed from requirements \cite{DBLP:journals/infsof/IrshadPP18}.
Such evaluation would make little sense, however: the SOOR approach just applies the object-oriented principles to the construction of requirements.
This makes the evaluation straightforward: the amount of duplication may be removed completely -- this is exactly what happens to software built around the same principles.
We prefer then to evaluate the extent to which the reuse approach simplifies specification of individual requirements.

Recall the ``Bounded Existence Between Q and R'' SRP (Equation \ref{eq:bounded_existence_between}).
Repeatedly instantiating and verifying this SRP as it is may be a challenging and error-prone process.
In our approach, the complexity of specifying a SOOR does not depend on the SOORT's internal complexity.
For example, a SOOR expressing requirement ``equinox happens not more than two times during a year'' for a calendar system would roughly look as follows:
\begin{center}
	\begin{lstlisting}
	class
	  EQUINOX_FREQUENCY
	inherit
	  BOUNDED_EXISTENCE_BETWEEN [CALENDAR, EQUINOX, YEAR_BEGINNING, YEAR_END]
	  CALENDAR _REQUIREMENT
	end
	\end{lstlisting}
\end{center}
where: class \e{CALENDAR} represents the specified type; \e{EQUINOX} captures the equinox condition; \e{YEAR_BEGINNING} and \e{YEAR_END} capture the beginning and the end of the year, respectively; \e{CALENDAR_REQUIREMENT} captures phenomena common to calendar requirements.
Consider now requirement ``the beginning of the year is always followed by the end of the year''.
This requirement represents the ``Global Response'' SRP, in LTL:
\begin{align*}
\square (P \implies \Diamond S)
\end{align*}
The complexity of this SRP is incomparably smaller than the complexity of the previous one.
The SOOR capturing the new requirement would look as follows:
\begin{center}
	\begin{lstlisting}
	class
      YEAR_END_RESPONDS_TO_YEAR_BEGINNING
	inherit
      RESPONSE_GLOBAL [CALENDAR, YEAR_END, YEAR_BEGINNING]
      CALENDAR _REQUIREMENT
	end	
	\end{lstlisting}
\end{center}
This SOOR is simpler only in one way: it provides 3 actual generic parameters to its SOORT, while the previous one provides 4.
We may say that the SOOR's complexity depends linearly on the number of formal generic parameters in the SOORT from which the SOOR inherits.
For the existing control software SRP's, however, this number never exceeds 4.

As for specifying SOORs for software components from the ADT SOORTs: the number of ADT axioms depends quadratically on the number of operations in the specified ADT \cite{DBLP:books/daglib/0025377}.
Specifying a SOOR from an ADT SOORT requires only to connect the deferred features of the SOORT with the concrete features of the specified type.
This does not remove the need to verify all the ADT axioms present in the SOORT in the form of specification drivers.
The AutoProof technology solves the verification problem.
The approach replaces the specification complexity from quadratic to linear.

%% file: sections/Summary.tex
\section{Conclusions} \label{summary:conclusions}

Our approach further removes the notational gap between software requirements and their implementations, following the key idea of DbC.
The approach makes software requirements:
\begin{itemize}
	\item \emph{Expressive}, thanks to the expressive power of an object-oriented programming language with contracts.
	\item \emph{Well-defined (when they take the traditional form of contracts)}, thanks to specification drivers and the powerful verification tool.
	\item \emph{Reusable}, thanks to the standard object-oriented techniques -- genericity and abstraction.
\end{itemize}
Quantitative (Section \ref{quantitative_evaluation}) arguments show the improvements in expressiveness, verifiability and reusability of the resulting requirements.

\section{Future work} \label{summary:future_work}

The following work is necessary to demonstrate the benefits of our approach in an even more convincing way:
\begin{itemize}
	\item Applying and measuring the approach in an industrial setting.
	\item Proving formally that the presented library of seamless requirement templates correctly resembles the encoded SRPs’ semantics.
\end{itemize}

The present work opens the following research directions:
\begin{itemize}
	\item Automatic generation of SOORTs in a given programming language from a given pattern expressed in some mathematical formalism.
	\item Extending the existing IDEs for better support of seamless requirements and their templates.
	\item AI-based detection of patterns in natural language requirements with their subsequent translation to SOORTs.
	\item Identifying new patterns in recurring requirements that do not map to existing patterns.
\end{itemize}